\documentclass[aip, jap, reprint, amssymb, amsfonts]{revtex4-1}

\usepackage{amsmath}
\usepackage{graphicx}
\usepackage[unicode]{hyperref}
\usepackage[latin1]{inputenc}
\begin{document}

\title{Observing ``Quantized'' Conductance Steps in Silver Sulfide: Two Parallel Resistive Switching Mechanisms}

\affiliation{Kamerlingh Onnes Laboratorium, Leiden University, PO Box 9504, 2300 RA Leiden, The Netherlands}
\author{Jelmer J.T. Wagenaar}
\author{Monica Morales-Masis}
\author{Jan M. van Ruitenbeek}
	\email{ruitenbeek@physics.leidenuniv.nl}
\noaffiliation

\begin{abstract}
We demonstrate that it is possible to distinguish two conductance switching mechanisms in silver sulfide devices at room temperature. Experiments were performed using a Ag$_2$S thin film deposited on a wide Ag bottom electrode, which was contacted by the Pt tip of a scanning tunneling microscope. By applying a positive voltage on the silver electrode the conductance is seen to switch to a state having three orders of magnitude higher conductance, which is related to the formation of a conductive path inside the Ag$_2$S thin film. We argue this to be composed of a metallic silver nanowire accompanied by a modification of the surrounding lattice structure. Metallic silver nanowires decaying after applying a negative voltage allow observing conductance steps in the breaking traces characteristic for atomic-scale contacts, while the lattice structure deformation is revealed by gradual and continuously decreasing conductance traces.
\end{abstract}

\maketitle

\section{Introduction}

It poses an interesting challenge to attempt to create atomic scale switches for implementation in future electronic devices, and some progress in this direction has been reported.\cite{Martin2009, Xie2004, Ter05, Ger11}  One of the proposed switches is based on cation migration and redox reactions in solid-state ionic conductors.\cite{Was07, Was09, Jam11} Terabe {\it et al.} \cite{Ter05} reported atomic switching behavior by electrochemical reactions taking place in a vacuum gap between micro fabricated Pt and Ag$_2$S electrodes. They showed switching between integer values of the unit of conductance, $G_0=(12.9\ \text{k}\Omega )^{-1}$, at room temperature and attributed this to the formation of a contact made up of a single atom, or a few atoms. Atomic conductance steps were also observed much earlier by Hajto et al.\cite{Haj91} in a metal/p$^+$-amorphous Si/metal thin film memory structures. \\


Here, we report on investigations on the conditions for observing atomic switching behavior without a vacuum gap.\cite{note_terabe} We used a device composed of a Ag$_2$S thin film deposited on top of a wide Ag layer. By contacting the thin film with the Pt tip of a scanning tunneling microscope (STM) and applying a positive bias voltage to the silver bottom layer the Ag$_2$S film was seen to switch to a high-conductive state. This high-conductive state is associated with the creation of a conducting path inside the Ag$_2$S film.\cite{Xu10} The Ag$_2$S film can be switched back to the low-conductive state by applying a negative bias voltage, which is associated with breaking up, or dissolution, of the conductive path. In analyzing 'on' to 'off' conductance traces we found evidence for the coexistence of two parallel breaking mechanisms.

\section{Experimental details}

 The deposition of Ag$_2$S was achieved by sputtering Ag in a Ar/H$_2$S plasma. It was deposited on top of a Ag film (200nm) which sits on a Si(100) substrate. The thickness of the silver sulfide layer is approximately 200nm and has a roughness of 30nm. The fabrication process and the characterization are described in more detail in a previous report by Morales {\it et al.}\cite{Mor10}

The measurements have been performed using a JEOL ultra-high vacuum STM (JSPM-4500A) at room temperature and at a pressure of $10^{-9}$mbar. The contact geometry is illustrated in the inset in Fig~\ref{figure:IVcurve}(a). A FEMTO (DLPCA-200) current amplifier was used to replace the standard JEOL current amplifier in order to cover a larger current amplification range. In terms of electron transport silver sulfide is a semiconductor, therefore control of the tip-sample distance requires biasing well below or well above the band gap.\cite{Mon11} In order to avoid early ion accumulation and the built up of a conducting path before the start of the experiment, the setting of the tip-sample distance requires a {\em negative} voltage on the sample, larger than the band gap. An external data acquisition card from National Instruments was added to the controller of the STM in order to apply the sample voltage and for measuring the current. The data acquisition card was controlled by a Labview program that was set up for measuring current-voltage (IV) characteristics, and traces of conductance (G) vs. time (t).

\section{Results}
\subsection{IV characteristics}
Before starting the measurements we need to confirm that the tip is in contact with our sample. This can be decided based upon the measured current-voltage relation. When a large tunneling gap is formed the resistance is dominated by vacuum tunneling and the IV curve is nearly linear, for sufficiently low bias. On the other hand, when in contact an exponential current-voltage characteristics is observed that is the result of ion accumulation near the contact at positive bias before full conductance switching occurs. The ions act as dopants and their accumulation results in an exponential increase of the electronic conductance. The IV curves are well described by the following expression,\cite{Mor10}

\begin{figure}[ttt]
\includegraphics[width=\columnwidth]{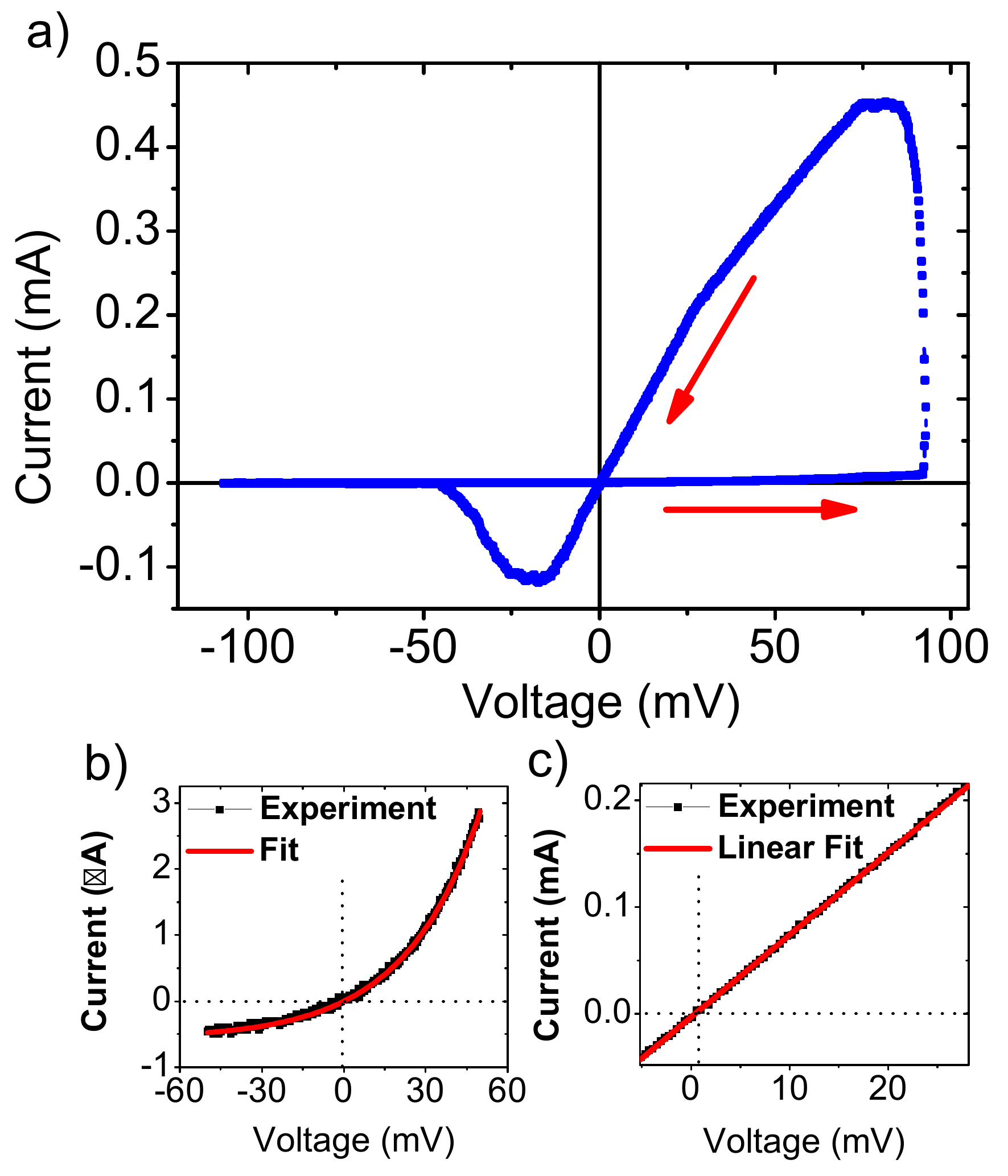}
\caption{ \textbf{a}) Current-voltage characteristics of a Ag-Ag$_2$S-Pt device (see inset) measured for a cycle duration of 1s. The ramp was started at 0V and follows the red arrows. In this IV curve full conductance switching is observed. The inset illustrates the contact geometry. \textbf{b}) Expanded scale view of the section of the IV curve in the 'off' state in (a). This part of the IV curve is fit very well by Eq.~(\protect\ref{equation:I}) (red curve), indicating that the Pt tip is in contact with the Ag$_2$S film, and that the film is in its equilibrium (semiconducting) state. The zero-bias conductance of the off-state is approximately 0.2G$_0$ for this contact size. \textbf{c}) On-state section of the IV curve in (a). At this stage the sample has a conductance of 100$G_0$, and the linear fit (red line) indicates metallic behavior.}
\label{figure:IVcurve}
\end{figure}

\begin{equation}
I(V)=K\sigma_0 \frac{k_BT}{e} \left(e^{(eV/k_BT)}-1\right)
\label{equation:I}
\end{equation}
with $\sigma_0 = 7.8\  10^{-2}\ \Omega^{-1} m^{-1}$ the electronic conductivity of Ag$_2$S at zero bias,\cite{Bon78} $T=295$ K is the temperature, $k_B$ is Boltzmann's constant, and $K$ is a geometrical factor with dimensions of length representing the contact size.

We measured IV curves by ramping the bias voltage and measuring the current with a sampling rate of 10000 samples per second. In fitting the data with Eq.~(\ref{equation:I}) $K$ is the only fitting parameter, from which we determine the size of the Pt STM tip contact.  Fig.~\ref{figure:IVcurve}(b) presents the IV curve for the low-conductance state (off-state) and the fit of the curve to Eq.~(\ref{equation:I}). From the quality of the fit we conclude that the sample is in its pristine, semiconducting, state and that the Pt tip is in contact with the sample. We can also conclude that there is no Joule heating of the sample since the temperature dependency is in the exponential. Increasing the voltage further causes switching to the on-state (Fig.~\ref{figure:IVcurve}(a)) due to the formation of a conductive path. Fig.~\ref{figure:IVcurve}(c) shows an IV curve for the on-state, and the linear fit indicates metallic behavior. Subsequently, returning to a sufficiently large negative voltage the sample switches back to the off-state.

\begin{figure}[bbb]
\centering
\includegraphics[width=\columnwidth]{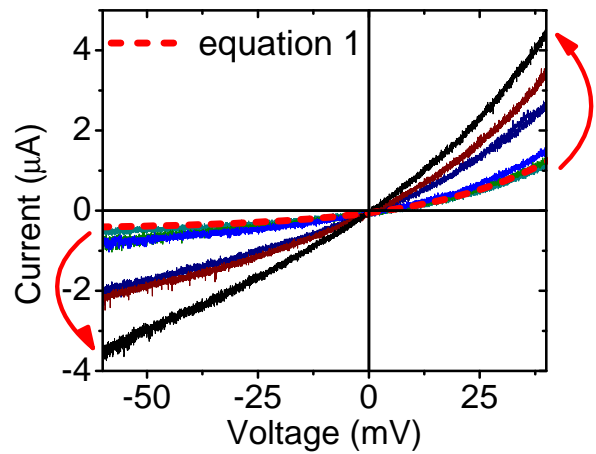}
\caption{Evolution of the off-state conductance curve for a Ag$_2$S film with an increasing number of switching cycles. Each of the IV curves in the plot is measured after a full switching cycle, such as the one shown in Fig.~\protect\ref{figure:IVcurve}a). The first curve (green) is measured after the first switching cycle, and the fit to Eq.~(\protect\ref{equation:I}) indicates that it remains close to the initial state of the pristine sample. The arrows give the direction of evolution for consecutive cycles. The zero-bias conductance of the junction changes from 0.1 $G_0$ (fitted curve) to approximately 1 $G_0$ (black curve).}
\label{figure:Change}
\end{figure}

After switching the device several times the off-state conductance of the sample becomes strongly modified. The evolution of the off-state IV curves with the number of switching cycles is presented in Fig.~\ref{figure:Change}. The switching cycles were similar to the one shown Fig.~\ref{figure:IVcurve} and the IV curves were recorded a few seconds after every cycle. Starting from the IV characteristics of the pristine sample (green curve) we observed the conductance increasing from 0.1$G_0$ to 1$G_0$ after the sixth cycle.

Initially, the IV curves are described well by Eq.~(\ref{equation:I}) (the red dashed curve). After several switching cycles the IV characteristic can only be fit by adding a significant linear term to Eq.~(\ref{equation:I}). This higher conductance state of the sample will eventually return to the initial conductance after applying a negative voltage for a longer period of time as will be shown below (Fig. \ref{figure:Mixed}). This changing sample conductance characteristics after switching has been reported previously and is referred to as the learning behavior of the switching mechanism.\cite{Has10, Ter07}, and is being explored for realizing artificial synapses.\cite{Ohn11} 

\subsection{Breaking}
 We measured traces of conductance as a function of time by control of the bias voltage in the following way: when the conductance was seen to fall below 0.5$G_0$ a positive bias voltage was applied to the sample. To achieve rapid switching we used a voltage of $+100$mV.\cite{Mon11} A high-conductance path was formed and the conductance was seen to rise to values above 100$G_0$. Once G was detected to pass above 100$G_0$ a negative bias voltage of $-100$mV was applied to the sample in order to break the conductive path, until the conductance approached the initial state of G $<$ 0.5$G_0$. In this way the formation and breaking process is more controllable than by applying fixed pulses to the sample.

\begin{figure}[tb]
\centering
\includegraphics[width=\columnwidth]{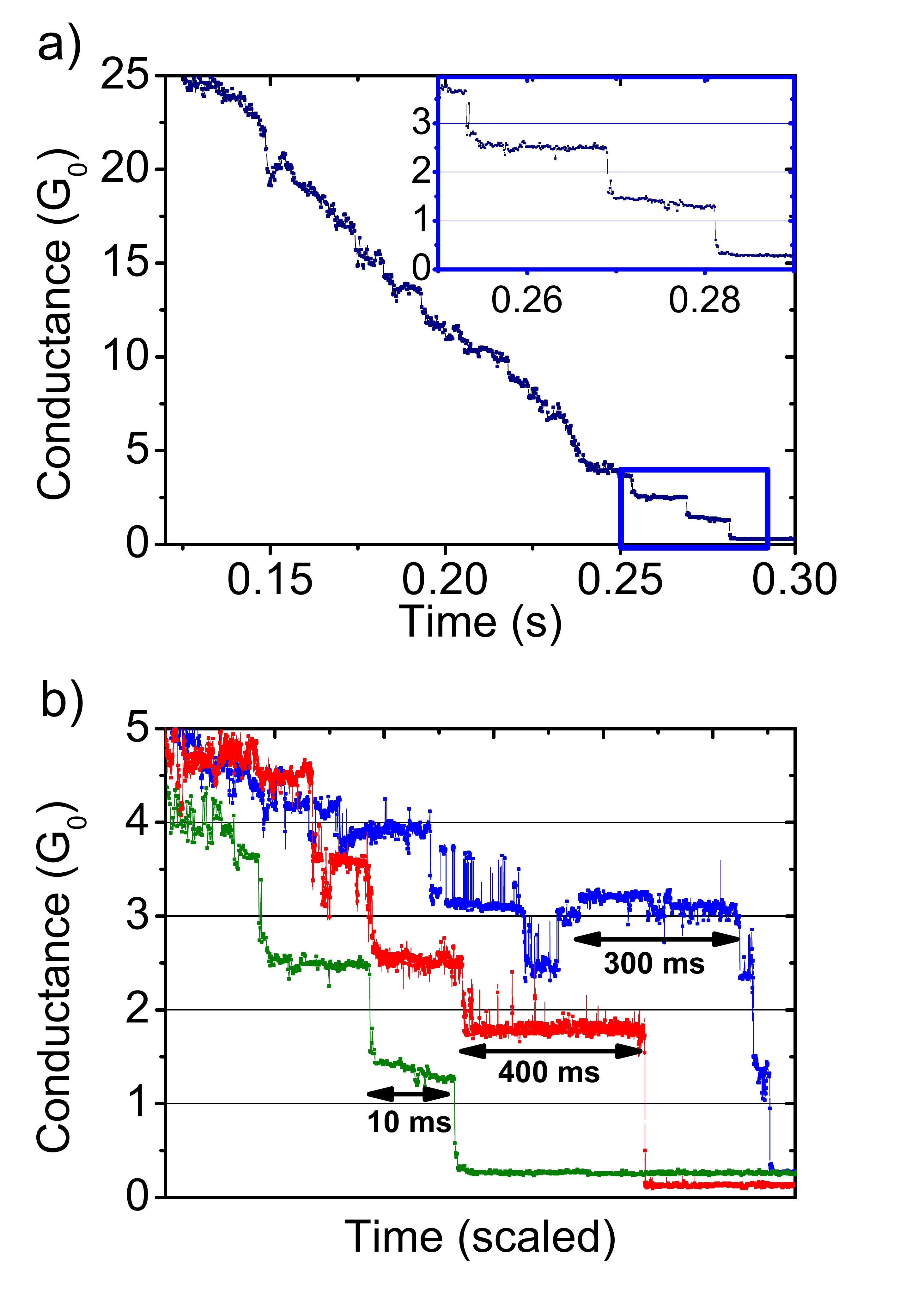}
\caption{Steps in the conductance become visible when breaking the conductive path at a bias of $-100$mV. The conductive path was formed by applying a voltage of $+100$mV, allowing the conductance to reach 100$G_0$ within a second. \textbf{a}) Breaking trace with three clear conductance steps of approximately 1$G_0$ observed at the last stages of breaking. The inset shows a zoom of the steps. \textbf{b}) Three breaking traces with atomic conductance steps having different lengths in time. The measurements were performed on different spots of the sample and using different Pt-tips. The upper trace shows two-level fluctuations that are typical for atomic size contacts, and are attributed to single atoms oscillating near the contact.}
\label{figure:Steps}
\end{figure}

We measured many breaking traces at different spots on the sample and we recognized two types of traces: traces having a step-like pattern, and traces showing only a slow and continuous decay of the conductance. The observed patterns in the traces indicates the coexistence of two switching mechanisms.

\begin{figure}[ttt]
\centering
\includegraphics[width=\columnwidth]{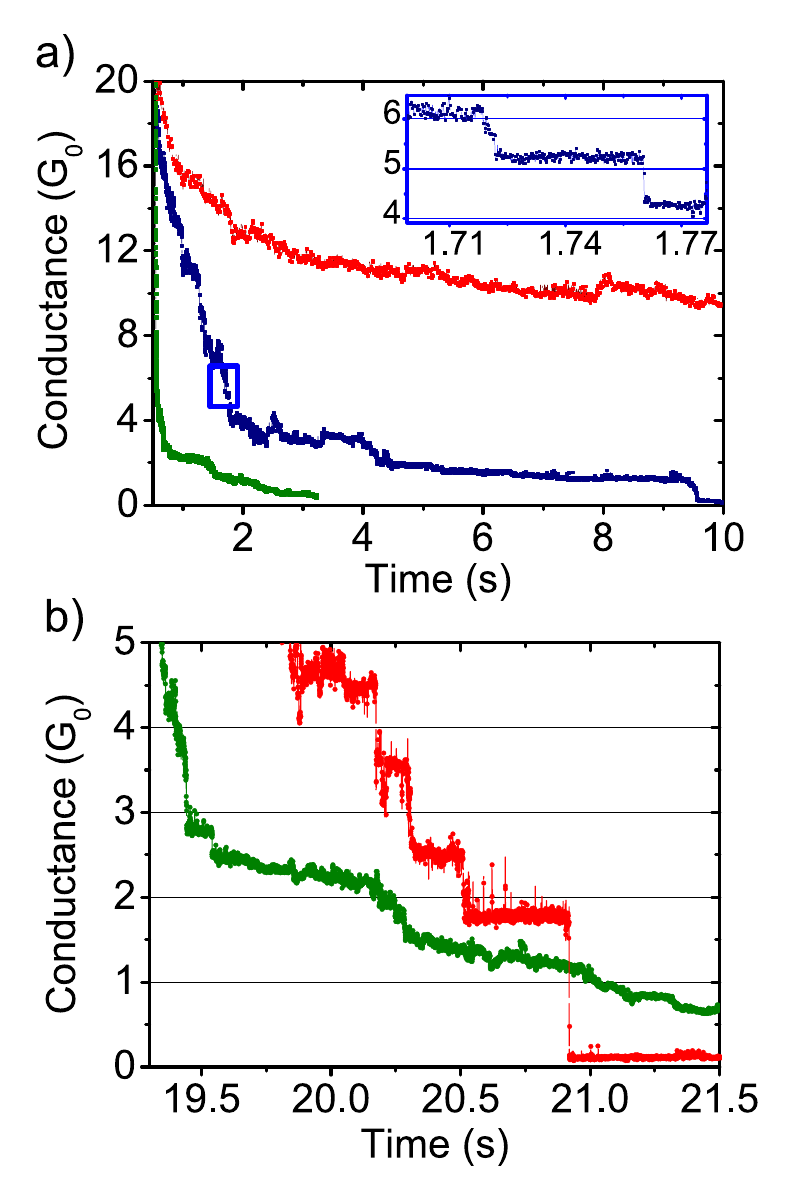}
\caption{\textbf{a})Continuous conductance changes observed when breaking a preformed conductive path at a bias voltage of $-100$mV. Mixed behavior is seen in the middle trace. The inset shows a magnification of the steps of approximately 1$G_0$. The upper trace shows only continuous behavior and illustrates the high-conductance that is visible in some measurements after switching several times. The lower trace also shows only continuous behavior but the high-conductance state decays in few seconds to a conductance around 0.1$G_0$. \textbf{b}) Comparison between a continuous trace (lower trace presented in (a)) and a trace with atomic conductance steps (red trace in Fig.~\ref{figure:Steps}(b)).}
\label{figure:Mixed}
\end{figure}

The first mechanism is the dissolution of a metallic silver conductance path showing, at the final stages of the breaking, atomic conductance steps. When a metallic silver filament dominates the conductance, upon applying a negative voltage the filament at its weakest point will become reduced to only a few bridging atoms. In this way quantum properties of the conductance of the silver filament will show up.\cite{Agr03} Fig.~\ref{figure:Steps} shows some examples of atomic conductance steps observed in conductance traces recorded as a function of the breaking time. In Fig.~\ref{figure:Steps}(a) one observes that the conductance decays almost linearly until arriving at about 5$G_0$, when steps of close to 1$G_0$ in height start to be visible. Very pronounced  plateaus and steps of approximately 1$G_0$ can also be seen in the plots of Fig.~\ref{figure:Steps}(b), with a very long plateau of 0.4s in the middle trace. The upper trace also shows two level fluctuations with an amplitude close to 1$G_0$ that is typical for atomic-scale contacts.\cite{Agr03}

\begin{figure}[bbb]
\centering
\includegraphics[width=\columnwidth]{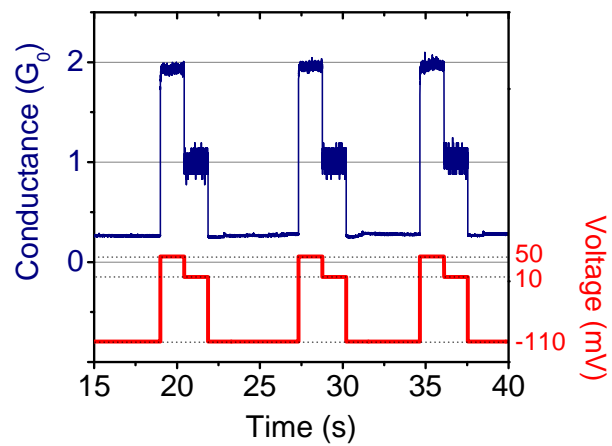}
\caption{Switching between targeted values of conductance by applying specific low-bias voltages to the Ag$_2$S device. In this experiment the voltages were chosen such as to obtain approximately the first two integer conductance values, in analogy to the experiments by Terabe {\it et al.}\cite{Ter05} Here we use fixed bias voltages instead of short pulses. This controllable switching can only be achieved after preparing the sample by several switching cycles. This example is chosen for illustration of the ambiguity that may arise when deciding whether a device is a true atomic scale switch.}
\label{figure:Control1}
\end{figure}

However, there appears to be a second mechanism active. We conclude this from the observation of a slower and nearly continuous decrease of the conductance. Traces with atomic conductance steps appeared only when the conductance rapidly dropped below 0.5$G_0$ in approximately one second. When the decrease in the conductance was slower, we observed behavior as illustrated by the upper trace in Fig.~\ref{figure:Mixed}. This is accompanied by a change in the IV characteristics similar to Fig.\ref{figure:Change}. We attribute this behavior to a second mechanism, most likely due to a modification of the local lattice structure giving rise to a region of increased conductance. This modification is probably induced by the electric field and the increased concentration of silver in the region of switching. It has been previously shown that the electric field can induced phase transitions, or decrease the phase transition temperature in materials like vanadium dioxide \cite{Per10} and complex perovskites \cite{Zha07}. After applying a negative voltage silver diffuses back to the Ag bottom reservoir and the lattice slowly relaxes to its equilibrium structure \cite{Xu10}.

Occasionally the two processes can be observed together, as illustrated in the lower trace in Fig.\ref{figure:Mixed}: around 6$G_0$ atomic conductance steps are visible, while somewhat later there is a continuous decrease over five seconds from $2$ to 1$G_0$. In terms of the two mechanisms described above this may be explained as being due to two parallel conductance paths.

\subsection{Controllable switching}
The continuous evolution of the IV characteristics in Fig.~\ref{figure:Change} and the continuous `on' to `off' conductance traces suggests that the local structure of the Ag$_2$S film has been modified. The conductance in this state can be controlled by applying positive bias voltages smaller than the threshold voltage. The voltages can be chosen to obtain specific values of conductance as illustrated in Fig.~\ref{figure:Control1}. In this example the conductance of the contact at a bias of -110mV was 0.3$G_0$, slightly higher than the conductance of the contact in the pristine state of 0.1 $G_0$.

This nanoscale resistive switch in the regime of the quantum of conductance should not be confused with an atomic scale switch. The steps that we attribute to intrinsic atomic scale structure (`quantization') are only short lived (Fig.\ref{figure:Steps}), and the IV curves in this state are linear, so that the conductance does not depend on the bias voltage. By controlling the second mechanism of conductance switching any conductance, including `quantized' values, can be set and maintained. The time scale for this process is much longer, and it allows manipulating the conductance over a wide range (Fig.~\ref{figure:Control1}).

\begin{figure}[ttt]
\centering
\includegraphics[width=5cm]{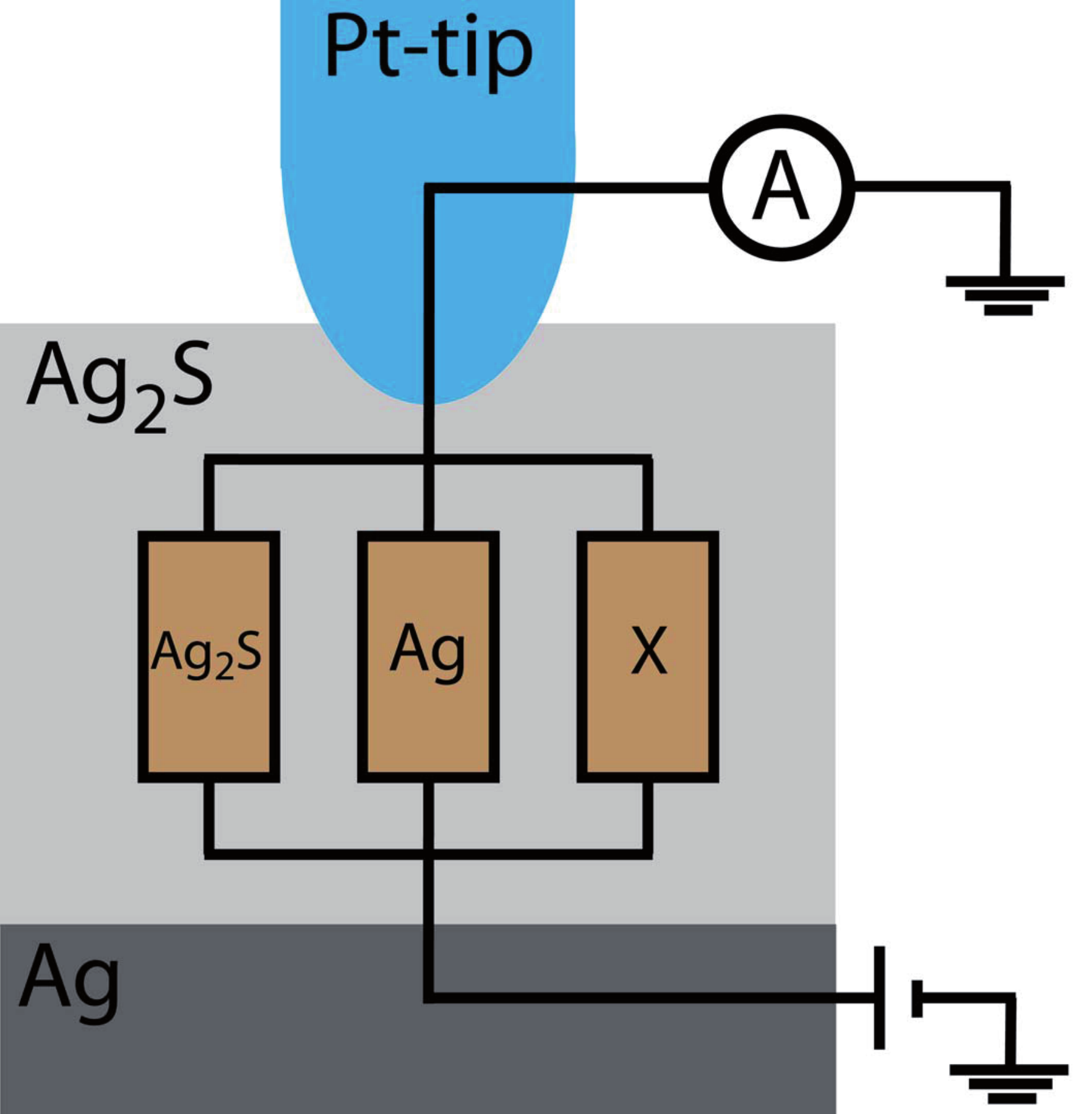}
\caption{The measurements suggest that three structures are involved in the conductance after full resistive switching. Before switching we have our pristine sample which obeys the IV relation of Eq.~(\protect\ref{equation:I}). During switching a high-conductive path is formed by metallic silver and, in parallel, by a modified structure of the silver sulfide (X). An interpretation of this modified structure comes from comparison of our results with Xu {\it et al.},\cite{Xu10} who identified it in HRTEM studies as the argentite phase of silver sulfide.}
\label{figure:Parallel}
\end{figure}

\section{Discussion}
We have observed two types of conductance breaking traces, which we associate to the occurrence of two switching mechanisms in silver sulfide. First, the presence of atomic conductance steps supports an interpretation in terms of the formation of a metallic silver filament. From previously performed experiments \cite{Agr03} we can state that a Ag atomic scale point contact presents steps in the conductance at the last stage before it breaks. However, the controllable switching, the continuously changing IV characteristics, and the gradually decreasing conductance traces cannot be explained by the formation of a metallic silver filament alone. The fact that the gradually decreasing conductance traces can take several seconds to return to the `off' conductance (Fig.~\ref{figure:Mixed}), and decay of concentration gradients in Ag$_2$S occurs much faster than a second \cite{Mor10} point towards the view that a modification of the lattice structure is induced, which we refer to as the second mechanism of conductance switching. Figure~\ref{figure:Parallel} shows a cartoon of the three different structures that may contribute in parallel to the total conductance: the pristine semiconducting sample (Ag$_2$S), metallic silver filaments (Ag), and the as yet undefined modified structure (X).\\

Our interpretation is consistent with the observations by Xu {\it et al.} \cite{Xu10} from \textit{in situ} switching measurements of a Ag/Ag$_2$S/W device. From real time measurements inside a high-resolution transmission electron microscope (HRTEM), Xu {\it et al.} determined that the conductive path formed when applying a positive voltage to the Ag electrode, is composed of a mixture of metallic Ag and argentite Ag$_2$S.\\

At room temperature, the equilibrium lattice structure of silver sulfide is the so called acanthite phase. Above 450 K, the lattice undergoes a phase transition to the argentite structure. Argentite has an electronic conductivity that is three orders of magnitude higher, and behaves like a metal.\cite{Kas03} The argentite structure has previously been stabilized at room temperature by rapidly cooling of silver sulfide from high temperatures.\cite{Kun06} According to Xu's interpretation this phase transition to the argentite phase is not caused by Joule heating because the currents are quite low at the off state, but is believed to be driven by the increased silver ion concentration and the applied field.\\

Adhering to this interpretation of the phase transition we explain the observed continuous traces and mixed traces shown in Fig.4 as follows. The bias voltage drives both switching mechanisms: metallic filament formation and the local partial phase transition. When a silver filament is formed that stretches fully across the thickness of the film its high conductance dominates the observed electron transport. Break down of the filament at the last stages produces a connection formed by just a few atoms, and when these disconnect one by one this becomes visible as near-quantized steps in the conductance. When the metallic silver of a filament dissolves very rapidly, or an incomplete filament is formed, atomic conductance steps will be absent and the conductance will drop continuously, as a result of the gradual decay of the locally modified structure. When the breaking of the silver filament occurs on the same time scale as the decay of the modified structure back to the initial room temperature phase, mixed behavior as seen in the middle curve in Fig.~\ref{figure:Mixed} can be observed. The conditions for the formation of a metallic silver filament and the observation of atomic conductance steps are not yet fully understood, since the two processes are controlled by the same bias voltage.

\section{Conclusion}
In summary, we identify two mechanisms of conductive path formation inside a thin film of silver sulfide. The first mechanism is the formation of a metallic silver filament which we associate with the observation of atomic conductance steps. Second, there is a modification of the silver sulfide structure to a higher-conductance phase. The conductance of this modified structure is continuously tunable, and allows setting the conductance to any target value by applying appropriate positive voltages. Contrary to previous reports,\cite{Ter07,Ohn11} we argue that this controllable switching cannot be attributed to an intrinsic atomic scale structure. The memory functions explored by Ohno {\it et al.}\cite{Ohn11} cannot be explained by a metallic filament formation alone. Our results point towards the role of the two switching mechanisms in deciding whether the information is maintained for shorter or longer times.

\begin{acknowledgments}
This work is part of the research program of the Dutch Foundation for Fundamental Research on Matter (FOM) that is financially supported by NWO.
\end{acknowledgments}


\begin{thebibliography}{21}
\bibitem [{not()}]{note_terabe}%
  \BibitemOpen
  \href@noop {} {}\bibinfo {note} {One may argue that the assumption made in
  Ref.~[\protect\onlinecite{Ter05}] of the formation of a vacuum gap prior to
  the switching operation is unlikely, and it is also not supported by
  evidence. For the functioning of the device it is not needed because
  switching occurs inside the Ag$_2$S material. Also, the initial state of the
  measurement in Ref.~[\protect\onlinecite{Ohn11}] is with the STM tip in
  contact with the silver sulfide, as was shown in
  Ref.~[\protect\onlinecite{Mon11}]}
\end{thebibliography}
\end{document}